\documentclass{article}
\usepackage{spconf,amsmath,amssymb,graphicx,booktabs,url}
\mathchardef\UrlBigBreakPenalty=100
\usepackage[T1]{fontenc}

\ninept
\newcommand{\bp}[1]{\textbf{\boldmath #1}}
\ifdefined{/Title (Text-Only LLM Correctors Break Correct Transcripts: A Repair-Damage Analysis of Audio Grounding in ASR Error Correction)
         /Author (Vasiliy Kudryavtsev, Kirill Borodin, Nikita Vasiliev, Mikhail Gorodnichev, Grach Mkrtchian)}\fi

\title{Text-Only LLM Correctors Break Correct Transcripts: A Repair--Damage\\
Analysis of Audio Grounding in ASR Error Correction}
\name{Vasiliy Kudryavtsev$^{1}$, Kirill Borodin$^{1,\ast}$, Nikita Vasiliev$^{1}$,
      Mikhail Gorodnichev$^{2}$, Grach Mkrtchian$^{3}$}

\address{$^{1}$BitmanagerAI, Dubai, UAE \qquad $^{2}$MTUCI, Moscow, Russia \qquad
$^{3}$lab260, Yerevan, Armenia\\[2pt]
{\normalsize\texttt{kborodin.research@gmail.com}}\\[2pt]
{\normalsize $^{\ast}$Corresponding author}}
\makeatletter
\def\@maketitle{\newpage
 \null
 \vskip 2em \begin{center}
 {\large \bf \@title \par} \vskip 1em {\large \lineskip .5em
\begin{tabular}[t]{c}\@name \\ \@address
 \end{tabular}\par} \end{center}
 \par
 \vskip 1em}
\makeatother
\usepackage{eso-pic}
\AddToShipoutPictureFG*{%
  \put(0,34){\makebox[\paperwidth]{\parbox{6.4in}{\centering\footnotesize
    This work has been submitted to the IEEE ICASSP for possible publication.
    Copyright may be transferred without notice, after which this version
    may no longer be accessible.}}}}

\begin{document}
\maketitle

\begin{abstract}
LLM correctors of ASR output are judged by aggregate WER, which hides what
they fix and what they break. We decompose every WER change exactly into
repaired tokens, damaged tokens and insertions, and use this to ask what the
source audio adds to a corrector. We compare correction with and without the
audio, for the model's own draft and for the output of an independent
recogniser, on three corpora. With Gemini 2.5 Flash, text-only correction of
its own draft breaks 13--32\% of already-correct transcripts; with the audio,
1--3\%. Damage falls in both settings. Repair falls against the own draft and
rises against the recogniser. On medical dictation, kept insertions cancel the
gain in net WER. The open-weight Voxtral, which rarely edits, does not show
this pattern. Prompts differ between the two arms, so the contrasts bound the
effect of the audio and do not isolate it.
\end{abstract}


\section{Introduction}
\label{sec:intro}

LLM post-correction of ASR output is now a common pipeline component with its
own survey \cite{peyghan2025survey}. It is also a component that can silently
rewrite a transcript \cite{koenecke2024careless}.

Whether it helps depends on the regime. Generative error correction (GER)
systems trained on $N$-best lists report gains
\cite{chen2023hyporadise,yang2023tap,yang2024gensec,hu2024robustger}.
Zero-shot prompting on a single hypothesis, the regime we study, mostly hurts.
Gu \emph{et al.} found that every LLM they tested degraded the ASR baseline
\cite{gu2024revisiting}. Ma \emph{et al.} found that zero-shot and one-shot
prompting raised the error rate \cite{ma2023can,ma2024asrec}. Wei \emph{et
al.} saw LLMs ``correct'' sentences that had no error \cite{wei2024asrec}.

One remedy is to give the corrector the audio as well as the text. Whispering
LLaMA and UADF fuse acoustic information into the corrector
\cite{radhakrishnan2023whispering,chen2024uadf}. ClozeGER feeds the speech to
a multimodal LLM because ``LLMs are unaware of the source speech during GER''
\cite{hu2024clozeger}. The ASR-EC benchmark finds this multi-modal setting the
best of the three it compares \cite{wei2024asrec}.

So grounding helps. What is missing is an account of what it changes. Every
result above is an aggregate error rate, and an aggregate rate mixes two
opposite flows: the corrector fixes some words and breaks others.

Counting the two flows separately is not new. It is standard for trained
correctors \cite{leng2021fastcorrect}, was recently proposed for multi-call
LLM protocols at the level of whole responses \cite{reitich2026correction},
and motivates confidence-gated correction \cite{hernandez2025confidence}. We
do it at the word level, where the split is exact with respect to WER, and ask
one question: what changes when the corrector can hear the audio?

On our primary model, Gemini 2.5 Flash, we find four things. (i) Text-only
correction breaks 13--32\% of the transcripts it receives in already-correct
form, and aggregate WER hides this. (ii) Grounding removes most of that
damage, whether the hypothesis is the model's own draft or the output of an
independent recogniser. (iii) The repair effect has opposite signs in these
two settings. (iv) On medical dictation, insertions hide the benefit of
grounding in net WER.

There are two limits. A near-ceiling synthetic corpus shrinks the repair
effect by an order of magnitude. An open-weight model that rarely edits shows
no consistent drop in damage.

Our reading is that grounding reduces damage for a corrector that
over-corrects and adds repair for one that under-corrects. With two models
this is a hypothesis. LLM self-correction of reasoning shows a similar
asymmetry: it fails without new external information
\cite{huang2024cannot,kamoi2024selfcorrection}.

Our contribution is the accounting, the factorial comparison and the findings
above.

\section{Method}
\label{sec:design}

\subsection{Conditions}

A corrector is a map $\hat{y} = g(a, h)$ from an audio signal $a$ and an input
hypothesis $h$ to an output transcript. Either input may be withheld.

We use two hypotheses. The \emph{ASR output} comes from a conventional
recogniser. The \emph{draft} is the single-pass transcript of the audio LLM
itself. A \emph{grounded} cell receives the audio; a \emph{text-only} cell
does not.

This gives five cells (Table~\ref{tab:factorial}). C and A correct the draft
without and with the audio: the \emph{self-correction} setting. T and F
correct the ASR output without and with the audio: the \emph{cross-system}
setting. B is a control, a second audio-only pass with no hypothesis; it
differs from the draft only in its prompt.

Every cell is scored against the system that supplied its input: C, A and B
against the draft, T and F against the ASR output. Within a setting the
hypothesis and the clips stay fixed and the audio is added or removed. The
prompt wording changes with it (Section~\ref{sec:prompts}).

\subsection{Repair, damage and insertions}

Let $r$ be the reference. We align $r$ with $h$ and with $\hat{y}$ and mark
every reference token as correctly recognised or not. \textsc{rep} counts the
tokens that are wrong in $h$ and right in $\hat{y}$; \textsc{dam} counts the
tokens that are right in $h$ and wrong in $\hat{y}$. Let $\iota_h$ and
$\iota_{\hat{y}}$ be the insertion counts and $W = |r|$, all summed over the
corpus. With $\Delta\mathrm{WER} = \mathrm{WER}(h) - \mathrm{WER}(\hat{y})$,
so that positive values are improvements,
\begin{equation}
\Delta\mathrm{WER} \;=\; \frac{\textsc{rep}}{W} - \frac{\textsc{dam}}{W} +
\frac{\iota_h - \iota_{\hat{y}}}{W}
\label{eq:identity}
\end{equation}
holds per utterance and in aggregate. We call the three summands the repair,
damage and insertion \emph{components} and write them rep, dam and ins. All
rates are in \% of reference tokens and all differences in percentage points
(pt). A positive ins means that the corrector removed insertions.

Equation~(\ref{eq:identity}) is an accounting identity, not a model. It holds
for any fixed alignment, but it has two blind spots. An edit that replaces one
wrong token with another registers in no component. And minimum-cost
alignments are not unique, so all three components depend on the path. We use
a deletion-first backtrace; an insertion-first implementation moved every
rate we report by at most $0.02$\,pt, a fifth of our smallest effect.

The \emph{break-rate} is the share of error-free inputs that acquire at least
one error. The \emph{fix-rate} is the share of erroneous inputs whose error
count decreases. The effect of the audio within a setting is the paired
difference between the grounded and the text-only cell (A$-$C or F$-$T).

\subsection{Prompts}
\label{sec:prompts}

The cells cannot share one wording: a text-only prompt has to say that there
is no audio. The Russian corpora use Russian versions of the English
templates. The grounded prompts are transcription tasks. They ask the model to
transcribe from the audio ``more carefully'' (B), using the hypothesis ``as a
hint'' (A) or ``as a PHONETIC GUIDE \ldots\ If ASR and audio clearly disagree
--- trust the audio'' (F), and not to add content absent from the audio.

The text-only prompts are editing tasks. C: ``Audio is unavailable \ldots\
Correct likely recognition mistakes only; do not add new facts \ldots\ If
unsure, leave the fragment unchanged.'' T: ``Fix errors: 1) Segmentation. 2)
Morphology. 3) Names and terms. 4) Do NOT delete or add words --- only fix
forms. 5) If a word looks unfamiliar but phonetically plausible --- keep it.''

The two arms of each contrast therefore differ in instructions as well as in
audio. For damage this works against our finding, because the text-only
prompts are the stricter ones. For repair it works in its favour: rules 4 and
5 of T forbid some edits that F is encouraged to make. The cross-system
repair effect is thus an upper bound on what the audio contributes.
Prompt effects can be large: B differs from the draft only in its prompt and
moves medical WER by $2.53$\,pt.

\section{Experimental setup}
\label{sec:setup}

\textbf{Corpora.} (i) A synthetic corpus of $3{,}000$ Russian utterances from
ten scientific domains, rendered with text-to-speech
\cite{rmndrnts_wikipedia_asr}. It is a near-ceiling diagnostic
(Section~\ref{sec:regime}). (ii) $300$ Russian audiobook utterances from
RuLibriSpeech \cite{bond005_rulibrispeech,borodin2025balalaika}. (iii) $300$
English medical dictation utterances \cite{ekacare_medical_asr}. Language is
confounded with corpus and recogniser. Requiring all seven outputs (ASR,
draft and five cells) to be non-empty leaves $n = 2{,}260$, $270$ and $284$.

\textbf{Systems.} The recogniser is GigaAM \cite{kutsakov2025gigaam} for
Russian and Whisper Large-v3-Turbo \cite{radford2023whisper} for English. The
primary corrector is Gemini 2.5 Flash \cite{comanici2025gemini25}. We repeat
all cells with the open-weight Voxtral-Small-24B \cite{liu2025voxtral}
($n = 467$, $96$, $96$ clips of the same corpora) and cell F with GPT Audio
(\texttt{openai/gpt-audio}; $n = 450$, $93$, $86$ clips with a usable output).
All models were called through one API provider between March and August
2026, with temperature $0$ and one call per utterance.

\textbf{Scoring.} We report corpus-level WER after NFKC normalization,
lower-casing, folding of the Cyrillic letter \emph{yo} onto \emph{ye}, and
punctuation stripping, with no inverse text normalization. Intervals are 95\%
paired utterance-level percentile bootstrap \cite{bisani2004bootstrap} with
$20{,}000$ resamples. They are unadjusted for multiplicity and ignore speaker
clustering, so we rely on sign consistency across corpora, not on single
borderline intervals.

\textbf{Exclusions.} Every comparison uses one common subset per corpus. For
Gemini the excluded utterances are missing outputs, not refusals. On the
synthetic corpus cell T has no output for 738 utterances and F for 269
(incomplete API runs); we did not verify that the common subset keeps the
domain balance. On the natural corpora 30 and 16 utterances drop out, almost
all because the draft is empty. That is a model failure, and excluding it
flatters the draft WER. Section~\ref{sec:regime} checks the rule.

\begin{table}[t]
\centering
\caption{Gemini 2.5 Flash. Top: WER (\%) of the two input systems. Below:
$\Delta\mathrm{WER}$ (pt) of each cell against the system that supplied its
input; positive is an improvement. Bold: 95\% interval excludes zero.}
\label{tab:factorial}
\vspace{3pt}
\setlength{\tabcolsep}{2.6pt}
\begin{tabular}{@{}lccc@{}}
\toprule
 & TTS RU & RuLibri RU & Medical EN \\
 & $n{=}2{,}260$ & $n{=}270$ & $n{=}284$ \\
\midrule
\multicolumn{4}{@{}l}{\emph{input systems (WER, \%)}}\\
ASR output (recogniser) & 3.95 & 7.29 & 16.06 \\
Draft (audio LLM) & 2.39 & 10.04 & 17.65 \\
\midrule
\multicolumn{4}{@{}l}{\emph{control, against the draft ($\Delta$WER, pt)}}\\
B\quad audio, no hypothesis & $+$0.01 & $+$0.18 & \bp{$+$2.53} \\
\multicolumn{4}{@{}l}{\emph{self-correction, against the draft}}\\
C\quad draft, no audio  & \bp{$-$2.78} & $+$0.38 & $+$0.88 \\
A\quad draft $+$ audio  & $+$0.14 & \bp{$+$0.83} & \bp{$+$0.79} \\
\multicolumn{4}{@{}l}{\emph{cross-system, against the ASR output}}\\
T\quad ASR output, no audio   & \bp{$+$0.69} & \bp{$+$0.48} & $-$0.05 \\
F\quad ASR output $+$ audio   & \bp{$+$2.17} & \bp{$+$1.65} & \bp{$+$1.52} \\
\bottomrule
\end{tabular}
\end{table}

\begin{table}[t]
\centering
\caption{Effect of the audio, Gemini 2.5 Flash: grounded minus text-only
cell, pt; net $=$ rep $-$ dam $+$ ins. Negative dam: less damage; positive
ins: fewer insertions left. Bold: 95\% interval excludes zero. Break-rates
(\%, text-only\,$\to$\,grounded) have no intervals.}
\label{tab:contrast}
\vspace{3pt}
\setlength{\tabcolsep}{2.6pt}
\begin{tabular}{@{}lccccc@{}}
\toprule
 & rep & dam & ins & net & break-rate \\
\midrule
\multicolumn{6}{@{}l}{\emph{self-correction: A (audio) vs.\ C (no audio)}}\\
TTS RU      & \bp{$-$0.15} & \bp{$-$2.83} & \bp{$+$0.24} & \bp{$+$2.92} & 31.9\,$\to$\,2.6 \\
RuLibri RU  & \bp{$-$0.50} & \bp{$-$0.88} & $+$0.08 & $+$0.45 & 16.3\,$\to$\,1.0 \\
Medical EN  & \bp{$-$0.74} & \bp{$-$1.72} & \bp{$-$1.08} & $-$0.09 & 12.8\,$\to$\,2.3 \\
\midrule
\multicolumn{6}{@{}l}{\emph{cross-system: F (audio) vs.\ T (no audio)}}\\
TTS RU      & \bp{$+$0.09} & \bp{$-$1.31} & \bp{$+$0.08} & \bp{$+$1.48} & 18.1\,$\to$\,2.6 \\
RuLibri RU  & \bp{$+$0.95} & $-$0.18 & $+$0.05 & \bp{$+$1.18} & \phantom{0}4.0\,$\to$\,3.0 \\
Medical EN  & \bp{$+$1.25} & \bp{$-$0.71} & $-$0.40 & \bp{$+$1.57} & 13.8\,$\to$\,6.9 \\
\bottomrule
\end{tabular}
\end{table}

\begin{table*}[t]
\centering
\caption{Illustrative examples from the medical corpus (Gemini 2.5 Flash,
normalized text); italics mark tokens changed by the corrector. Rows 1--3:
self-correction, the input is the draft; row 2 shows the first tokens of a
32-token utterance. Rows 4--5: cross-system, the input is the Whisper output.}
\label{tab:examples}
\vspace{3pt}
\begin{tabular}{@{}rllll@{}}
\toprule
\# & Reference & Input hypothesis & Output without audio & Output with audio \\
\midrule
1 & loss of hair          & loss of hair          & loss of \emph{hearing}  & loss of hair \\
2 & postoperative imaging \ldots & post operative imaging \ldots & \emph{postoperative} imaging \ldots & post operative imaging \ldots \\
3 & forxiga 10mg tablet   & forxa 10 mg tablet    & \emph{forxiga} 10 mg tablet & forxa 10 mg tablet \\
4 & seizure               & see you soon          & see you soon            & \emph{seizure} \\
5 & oxiconazole           & oxyconosol            & \emph{oxycontin}        & \emph{oxiconazole} \\
\bottomrule
\end{tabular}
\end{table*}

\section{Results}
\label{sec:results}

\subsection{Net effect}

Table~\ref{tab:factorial} shows the five cells. In the cross-system setting
the grounded cell beats the text-only cell on every corpus (F $>$ T, all
intervals excluding zero).

Self-correction is less clear. A beats C on the synthetic corpus
($+2.92$\,pt). On the natural corpora we cannot detect a net difference
($+0.45$\,pt, interval $[-0.13, +1.05]$; $-0.09$\,pt, $[-1.02, +0.82]$); the
intervals allow about one point either way. On medical speech the interval of
A excludes zero and that of C does not, but the two estimates do not differ.

By net WER alone, the audio does nothing for self-correction on natural
speech. The components tell a different story.

\subsection{Repair, damage and break-rate}

Table~\ref{tab:contrast} separates the flows. In self-correction the grounded
cell has lower repair \emph{and} lower damage on all three corpora, and all
six intervals exclude zero. Repair falls by $0.15$--$0.74$\,pt and damage by
$0.88$--$2.83$\,pt. With the audio the model simply changes fewer tokens. We
call this restraint, as a description and not as a mechanism.

In the cross-system setting damage also falls on all three corpora, with
intervals excluding zero on two. But repair \emph{rises} on all three
($+0.09$, $+0.95$, $+1.25$\,pt). On medical dictation it goes from $2.06$\% to
$3.31$\% of reference tokens, a 61\% relative increase. Part of this rise may
come from the prompts (Section~\ref{sec:prompts}).

The break-rate shows the restraint directly. In self-correction the text-only
cell breaks 31.9\%, 16.3\% and 12.8\% of the error-free drafts it receives,
although its prompt is the stricter one. The grounded cell breaks 2.6\%,
1.0\% and 2.3\%. In the cross-system setting the text-only cell breaks 18.1\%,
4.0\% and 13.8\%, and the drop is smaller: to 2.6\%, 3.0\% and 6.9\%.

On the natural corpora these rates rest on few error-free inputs, carry no
intervals and include orthographic-only changes, so we read them
descriptively. None of this is visible in aggregate WER there.

Table~\ref{tab:examples} shows illustrative cases. Without the audio the model
turns the correct draft ``loss of hair'' into ``loss of hearing''; with the
audio it leaves it alone. In the cross-system setting the grounded cell turns
``see you soon'' into the correct ``seizure'', an edit that the text-only
prompt does not allow. In row~5 the text-only cell swaps one wrong drug name
for another, which the identity does not register.

\subsection{Insertions}
\label{sec:ins}

Why does restraint not lower WER on medical dictation? Because of insertions.
The draft contains insertions worth $4.47$\% of reference tokens. The
text-only cell removes a quarter of them (to $3.29$\%). The grounded cell
leaves the total almost unchanged ($4.37$\%). The resulting $-1.08$\,pt
(interval excluding zero) cancels the $+0.98$\,pt of the repair and damage
components and gives the net $-0.09$\,pt.

On RuLibriSpeech the insertion component is small ($+0.08$\,pt). There the
lost repair offsets much of the reduced damage, and the interval is too wide
to decide.

What are these insertions? Of the 182 inserted tokens in the medical draft,
77 are halves of compounds that the reference writes as one token (``post
operative'', ``10 mg''). After the text-only cell 40 such tokens remain. After
the grounded cell 80 remain, and it also adds fillers it hears (``uh'': 6 in
the draft, 14 after grounding).

Each join of a split compound also counts as one repaired token. With no
inverse text normalization, all three components on this corpus are therefore
sensitive to the spelling conventions of the reference, and a compound- and
filler-insensitive scoring could flip the sign of the net A$-$C contrast.

Rows~2 and~3 of Table~\ref{tab:examples} show both sides. The text-only cell
joins ``post operative'' and corrects a drug name from its own knowledge. The
grounded cell does neither, and the second is a genuine repair that grounding
loses.

\subsection{Robustness and the synthetic-corpus ceiling}
\label{sec:regime}

We recomputed each contrast on the utterances complete for just its own two
cells. No component of Table~\ref{tab:contrast} crosses zero and no interval
changes side. We also scored every missing output as if the corrector had returned its
input unchanged. This sets repair and damage to zero on the imputed rows, so
it is a bound and not a neutral check: the synthetic F$-$T repair becomes
$+0.31$ instead of $+0.09$\,pt and damage $-0.98$ instead of $-1.31$\,pt.
Signs and intervals do not depend on the exclusion rule; levels do.

The synthetic corpus sits near its ceiling: 65\% of ASR outputs and 79\% of
drafts are already exactly correct, so damage is almost the only thing that
can happen. Three of its conclusions do not hold on natural speech.

Text-only self-correction looks very harmful there ($-2.78$\,pt), while on
the natural corpora its interval includes zero. A second pass without a
hypothesis looks useless or mildly harmful there ($+0.01$\,pt on the common
subset, $-0.38$\,pt over all outputs, Table~\ref{tab:support}) but is the
largest single gain on medical dictation ($+2.53$\,pt). And the cross-system
repair effect is an order of magnitude smaller there ($+0.09$ against $+0.95$
and $+1.25$\,pt).

\subsection{Domain terminology, per-domain results and noise}
\label{sec:support}

The measurements of Table~\ref{tab:support} exist for the synthetic corpus
only and carry the caveat above. Each cell is scored over its own complete
outputs, so columns are not paired and levels differ slightly from
Table~\ref{tab:factorial}.

Domain-term recall (DTR) is the share of the corpus's annotated domain terms
that appear in the output. Text-only self-correction lowers it from $69.0$\%
to $66.1$\%, close to the recogniser ($65.5$\%). The grounded cell keeps it
($69.3$\%). In the cross-system setting both cells raise DTR, the grounded one
twice as much ($+4.2$ against $+2.0$\,pt).

T lowers the WER of the ASR output but raises its CER, and C almost quadruples
the CER of the draft. This is consistent with whole words being replaced, not
respelled: without the audio, damage lands on the rare domain words. C raises
the WER of its draft in each of the ten domains (by $1.7$ to $4.0$\,pt), and A
is better than C in each.

The last rows repeat cells C, A and B on 250 clips, in their original form and
with reverberation (RT60 $0.5$\,s) plus pink-and-babble noise at 15 and 8\,dB
SNR. The WER of A almost triples. The absolute gap between C and A stays at
$2.84$, $2.96$ and $2.89$\,pt, while the relative gap narrows (C/A from $2.3$
to $1.5$). We did not decompose the noisy runs, so this shows that the WER gap
persists, not that the restraint does. The cost of withholding the hypothesis
(B against A) grows from $0.57$ to $1.43$\,pt.

\begin{table}[t]
\centering
\caption{Synthetic corpus, Gemini 2.5 Flash, raw rates in \% (WER and CER:
lower is better; not $\Delta$WER). Top: each system over its own complete
outputs ($n = 2{,}262$--$3{,}000$); columns cover different utterance sets and
are not paired. Bottom: WER on 250 clips, original and with reverberation and
noise. No intervals are available.}
\label{tab:support}
\vspace{3pt}
\setlength{\tabcolsep}{2.8pt}
\begin{tabular}{@{}lccccccc@{}}
\toprule
 & ASR & T & F & draft & C & A & B \\
\midrule
WER & 3.79 & 3.26 & 1.74 & 2.37 & 5.12 & 2.18 & 2.75 \\
CER & 0.79 & 1.23 & 0.42 & 0.67 & 2.57 & 0.67 & 1.18 \\
DTR & 65.5 & 67.5 & 69.7 & 69.0 & 66.1 & 69.3 & 69.1 \\
\midrule
WER, original & -- & -- & -- & -- & 5.03 & 2.19 & 2.76 \\
WER, 15\,dB   & -- & -- & -- & -- & 6.16 & 3.20 & 4.05 \\
WER, 8\,dB    & -- & -- & -- & -- & 8.91 & 6.02 & 7.45 \\
\bottomrule
\end{tabular}
\end{table}

\subsection{Other models}
\label{sec:voxtral}

Table~\ref{tab:voxtral} repeats all cells with Voxtral-Small-24B on a
659-utterance subset of the same clips. Its text-only self-correction rarely
edits: the fix-rate of cell C is 1.5--3.8\%, against 24.2--27.8\% for Gemini
(not tabulated).

Here the pattern of Gemini does not replicate. The grounded cell has
\emph{higher} repair in all six contrasts ($+0.24$ to $+4.64$\,pt), four with
intervals excluding zero, and repair rises even against Voxtral's own draft.
Damage moves inconsistently: it falls on medical dictation in both settings
and rises elsewhere. The insertion component, obtained by subtraction, is the
largest of the three in self-correction and makes the net A$-$C contrast
negative on two of three corpora.

In our two models, then, restraint is not a property of the acoustic channel
as such. It is consistent with what grounding does to a corrector that would
otherwise over-correct.

GPT Audio calls for the same caution. In cell F it degrades two of three
corpora (by $8.82$ and $8.54$\,pt) with break-rates of 21.3\% and 31.4\%. These
figures carry no intervals and cover only clips with a usable output.

Cell B suggests a second role of the hypothesis. On medical dictation Voxtral
without a hypothesis goes from a WER of $43.86$\% to $65.01$\%, and $99.6$\% of
that degradation is the insertion component: it emits $0.50$ inserted words
per reference token. Here the hypothesis bounded the output length, but the interval
includes zero and the opposite holds on RuLibriSpeech ($+6.09$\,pt).

\begin{table}[t]
\centering
\caption{Voxtral-Small-24B on subsets of the same clips; layout and units as
in Tables~\ref{tab:factorial} and~\ref{tab:contrast}. The ins rows are
obtained by subtraction and have no intervals. Bold: 95\% interval excludes
zero.}
\label{tab:voxtral}
\vspace{3pt}
\setlength{\tabcolsep}{2.6pt}
\begin{tabular}{@{}lccc@{}}
\toprule
 & TTS RU & RuLibri RU & Medical EN \\
 & $n{=}467$ & $n{=}96$ & $n{=}96$ \\
\midrule
\multicolumn{4}{@{}l}{\emph{input systems (WER, \%)}}\\
ASR output (recogniser) & 5.17 & 8.02 & 17.03 \\
Draft (audio LLM) & 7.90 & 26.31 & 43.86 \\
\midrule
\multicolumn{4}{@{}l}{\emph{cells ($\Delta$WER, pt)}}\\
B\quad audio, no hypothesis & $-$1.61 & $+$6.09 & $-$21.15 \\
C\quad draft, no audio  & \bp{$-$0.38} & $+$0.06 & $+$0.62 \\
A\quad draft $+$ audio  & $-$0.97 & \bp{$+$9.58} & $-$2.41 \\
T\quad ASR output, no audio   & $-$1.55 & $-$0.44 & $-$0.86 \\
F\quad ASR output $+$ audio   & \bp{$+$2.14} & $-$0.68 & $+$1.01 \\
\midrule
\multicolumn{4}{@{}l}{\emph{self-correction: A (audio) vs.\ C (no audio), pt}}\\
rep & \bp{$+$0.71} & \bp{$+$4.64} & $+$0.55 \\
dam & $-$0.15 & \bp{$+$0.68} & \bp{$-$1.55} \\
ins & $-$1.45 & $+$5.56 & $-$5.13 \\
\multicolumn{4}{@{}l}{\emph{cross-system: F (audio) vs.\ T (no audio), pt}}\\
rep & \bp{$+$2.46} & \bp{$+$0.75} & $+$0.24 \\
dam & \bp{$+$0.56} & \bp{$+$1.43} & \bp{$-$0.93} \\
ins & $+$1.79 & $+$0.44 & $+$0.70 \\
\bottomrule
\end{tabular}
\end{table}

\section{Discussion and limitations}
\label{sec:disc}

\textbf{Interpretation.} Why would the repair sign differ? One reading: the
draft was produced from the same audio, so hearing it again tells the model
nothing new, whereas for the ASR output the audio is new evidence. LLM
self-correction likewise needs external information \cite{huang2024cannot}.

This is a hypothesis. The two settings also differ in input WER, error type,
prompt and recogniser, and we did not test the interaction. The sign
difference holds between the two settings we ran; it is not an established
effect of who wrote the hypothesis. Voxtral contradicts the reading.

Nor can the data exclude that cell A is a near-deterministic re-transcription
that ignores the draft. At temperature $0$ that alone would give a low
break-rate and low repair. On medical dictation B, which has no draft, beats
A ($+2.53$ against $+0.79$\,pt), so the draft may anchor A to its own errors.

\textbf{Practical points.} On this model, extra repair from the audio appeared
only with an independent hypothesis. In self-correction its measurable benefit
on natural speech is fewer broken utterances, not lower WER. None of this
transfers to a model that rarely edits.

\textbf{Limitations.} Within a setting the hypothesis and the clips are fixed,
but the instructions change with the audio (Section~\ref{sec:prompts}); only
a matched-prompt design would isolate the audio. Only one model carries the
full factorial, and two of the three audio LLMs are commercial APIs subject to
version drift. The synthetic corpus is derived from Wikipedia text that the
LLMs have probably seen in pretraining; cf.\ the contamination-aware protocol
of \cite{hao2026frisian}. We do not compare against a corrector trained for
the task \cite{gu2024revisiting}.

\section{Conclusion}
\label{sec:concl}

Aggregate WER hides both the damage done by text-only LLM correction and what
the audio changes. We recommend reporting the break-rate and the three components next to
WER and, for a corrector that over-corrects, spending the audio call where the
hypothesis comes from an independent recogniser.

\section{Acknowledgments}
\label{sec:ack}

No funding was received for conducting this study. The authors have no
relevant financial or nonfinancial interests to disclose. Claude Fable 5.1
(Anthropic) was used only to edit and polish the text; the experiments, data,
analysis and conclusions are the authors' own, and the authors reviewed all
text.

\section{Compliance with Ethical Standards}
\label{sec:ethics}

This is a numerical study on publicly released speech corpora: Wikipedia text
rendered with text-to-speech, public-domain audiobook recordings, and a public
medical ASR evaluation set with no protected health information. No new
human-subject data was collected and no ethical approval was required.

\let\origthebibliography\thebibliography
\def\thebibliography#1{\origthebibliography{#1}\setlength{\itemsep}{0.15ex}\setlength{\parskip}{0pt}}
\bibliographystyle{IEEEbib}
\bibliography{refs}

\end{document}